\documentclass[journal]{IEEEtran}
\usepackage{amsmath}
\usepackage{breqn}
\usepackage{graphicx}
\usepackage{subcaption}
\usepackage{mwe}
\usepackage{cite}

\title{Evaluating Measurement-Based Dynamic Load Modeling Techniques and Metrics}

\author{Phylicia~Cicilio,~\IEEEmembership{Student Member,~IEEE,}
        and~Eduardo~Cotilla-Sanchez,~\IEEEmembership{Senior Member,~IEEE}
\thanks{This material is based upon work supported by the National Science Foundation Graduate Research Fellowship under Grant No.~1314109-DGE.}
\thanks{The authors are with the School of Electrical Engineering and Computer
Science, Oregon State University, Corvallis, OR 97331 USA e-mail: (ciciliop@oregonstate.edu; ecs@oregonstate.edu).}}

\date{}

\begin{document}
\maketitle
\begin{abstract}
Wide-area data and algorithms in large power systems are creating new opportunities for implementation of measurement-based dynamic load modeling techniques. These techniques improve the accuracy of dynamic load models, which are an integral part of transient stability analysis. Measurement-based load modeling techniques commonly assume response error is correlated to system or model accuracy. Response error is the difference between simulation output and phasor measurement units (PMUs) samples.  This paper investigates similarity measures, output types, simulation time spans, and disturbance types used to generate response error and the correlation of the response error to system accuracy.  This paper aims to address two hypotheses:~1) can response error determine the total system accuracy? and 2) can response error indicate if a dynamic load model being used at a bus is sufficiently accurate? The results of the study show only specific combinations of metrics yield statistically significant correlations, and there is a lack of pattern of combinations of metrics that deliver significant correlations. Less than 20\% of all simulated tests in this study resulted in statistically significant correlations. These outcomes highlight concerns with common measurement-based load modeling techniques, raising awareness to the importance of careful selection and validation of similarity measures and response output metrics. Naive or untested selection of metrics can deliver inaccurate and misleading results. 

\end{abstract}

\section{Introduction}

The introduction of phasor measurement units (PMUs) and advanced metering infrastructure (AMI) has ushered in the era of big data to electrical utilities. The ability to capture high-resolution data from the electrical grid during disturbances enables the more widespread use of measurement-based estimation techniques for validation of dynamic models such as loads. Transient stability studies use dynamic load models. These studies are key for ensuring electrical grid reliability and are leveraged for planning and operation purposes \cite{Shetye}. It is imperative that dynamic load models be as representative of the load behavior as possible to ensure that transient stability study results are accurate and useful. However, developing dynamic load models is challenging, as they attempt to represent uncertain and changing physical and human systems in an aggregate model. 

Several methods exist for determining load model parameters, such as measurement-based techniques using power systems sensor data \cite{Kim,Kontis,Zhang,Renmu,Choi_2,Ma_2}, and methods that use parameter sensitivities and trajectory sensitivities \cite{Choi,Kim,Son,Ma,Zhang}. A common practice in measurement-based techniques is to use system response outputs, such as bus voltage magnitude, from PMU data and simulation data and compare the output with a similarity measure, such as Euclidean distance. The error between PMU data and simulation output is referred to as response error in this paper. As investigated in \cite{siming_thesis}, the underlying assumption that reducing response error results in a more accurate model and system is not guaranteed. 

This paper examines the relationship between response error and system and model accuracy to highlight concerns with common measurement-based technique practices. The methods used in the study examine whether the selection of a load model is accurate at a given bus. Measurement-based techniques typically perform dynamic load model parameter tuning to improve accuracy. In parameter tuning, significant inter-dependencies and sensitivities exist between many dynamic load model parameters \cite{Choi,Kim,Son,Ma,Zhang}, which is one of the reasons why dynamic load model parameter tuning is challenging. This study compares the selection of two loads models, the dynamic composite load model (CLM) and the static ZIP model instead of parameter tuning. The static ZIP model is the default load model chosen by power system simulators and represents loads with constant impedance, current, and power. The CLM load model has become an industry standard, particularly for the western United States, which represents aggregate loads including induction machine motor models, the ZIP model, and power electronics \cite{Kosterev,Renmu}. The choice of changing the load model is made to compare known differences in responses from load motor models with the CLM model and static load models with the ZIP model. By comparing load model selection, the presence of a correlation between response error and system accuracy will be assessed.  

This study performs two experiments to address two main hypotheses. The first experiment is a system level experiment to test hypothesis 1) can response error determine the total system accuracy of how many load models at buses in the system are accurate? The second experiment is a bus level experiment to test hypothesis 2) can response error indicate if a load model being used at a bus is accurate? The results from these experiments demonstrate that it can't be assumed that response error and system accuracy are correlated. The main contribution of this paper is to identify the need for validation of techniques and metrics used in dynamic load modeling, as frequently used metrics can deliver inaccurate and meaningless results. 

The remainder of this paper is organized as follows. Section II discusses the use of dynamic load models in industry and those used in this paper. In Section III, similarity measures are discussed in relevance to power systems time series data. Section IV details the methodology used to evaluate the system level experiment of hypothesis 1. Section V provides and discusses the results from system level experiment. Section VI details the methodology used to evaluate the bus level experiment of hypothesis 2. These results are provided and discussed in Section VII. In conclusion, Section VIII discusses the implication of the results found in this study and calls for attention to the importance of careful selection and validation of measurement-based technique metrics. 

\section{Similarity Measures}\label{similarity_section}
A similarity measure compares how similar data objects, such as time series vectors, are to each other. A key component of measurement-based techniques is to use a similarity measure to calculate the response error. Then typically, an optimization or machine learning algorithm reduces this response error to improve the models or parameters in the system. Several measurement-based dynamic load model estimation studies employ Euclidean distance as a similarity measure \cite{Renmu,Visconti,Kong} . However, there are characteristics of power systems time series data which should be ignored or not emphasized, such as noise, which are instead captured by Euclidean distance. Power system time series data characteristics include noise, initialization differences, and oscillations at different frequencies. These characteristics result in shifts and stretches in output amplitude and time as detailed in Table \ref{similarity_measures}. 

\begin{table}[!t]
\renewcommand{\arraystretch}{1.3}
\caption{Examples of amplitude and time shifting and stretching \cite{siming_thesis}}
\label{similarity_measures}
\centering
\begin{tabular}{p{1.2cm}| p{2.8cm} p{2.8cm}}
\hline
 & \bfseries Amplitude & \bfseries Time\\
\hline
\bfseries Shift & initialization differences, discontinuities & different/unknown initialization time\\
\bfseries Stretch & noise & oscillations at different frequencies  \\
\hline
\end{tabular}
\end{table}

The characteristics listed in Table \ref{similarity_measures} are the effect of specific phenomena in the system. For example, differences in control parameters in motor models and potentially also playback between motor models can cause oscillations at different frequencies. Certain changes in output are important to capture as they have reliability consequences to utilities. An increase in the initial voltage swing after a disturbance can trip protection equipment. An increase in the time it takes for the frequency to cross or return to 60 Hz in the United States has regulatory consequences resulting in fines. Response error produced by similarity measures should capture these important changes. Other changes to output, such as noise, should be ignored.

Different situations when comparing simulation data to simulation data versus comparing simulation data to PMU data cause some characteristics listed in Table \ref{similarity_measures}. Comparing simulation data to simulation data occurs in theoretical studies, and comparing simulation data to PMU data would be the application for utilities. Initialization differences and differences in initialization time can occur when comparing simulation data to PMU data due to the difficulty in perfectly matching steady-state values. However, when comparing simulation data to simulation data, initialization differences and differences in initialization time likely highlight errors in the simulation models, parameters, or values.  

Similarity measures have the capability to be invariant to time shift and stretch or amplitude shift and stretch. Table \ref{similarity_measures2} lists the similarity measures examined in this study with their corresponding capabilities. These similarity measures are chosen to test the sensitivities to all four quadrants of Table \ref{similarity_measures}. 

\begin{table}[thb]
\centering
\caption{\label{similarity_measures2} Similarity measures capabilities} 
\label{similarity_measures2}
\begin{tabular}{c| p{1.1cm} p{1.1cm} p{0.9cm} p{0.9cm} }
\hline \\
& \centering \bfseries Amplitude Shift & \centering \bfseries Amplitude Stretch & \centering \bfseries Time Shift & \bfseries Time Stretch  \\

\hline
\bfseries Euclidean Distance &  &  &  &  \\
\bfseries Manhattan Distance &  &  &  &  \\
\bfseries Dynamic Time Warping &  &  & \centering $\bullet$  & $\bullet$ \\
\bfseries Cosine Distance & \centering $\bullet$  &  &  & \\
\bfseries Correlation Coefficient & \centering $\bullet$ & \centering $\bullet$ &  & \\
\hline
\end{tabular}
\end{table}

Euclidean distance and Manhattan distance are norm-based measures which are variant to time and amplitude shifting and stretching. Euclidean distance is one of the most commonly used similarity measures in measurement-based techniques. These norm based distances can range from 0 to $\infty$. \

The cosine similarity takes the cosine of the angle between the two vectors to determine the similarity. By only using the angle between the vectors, this similarity is invariant to amplitude shifting \cite{siming_thesis}. This similarity can range from -1 to 1.

The Pearson correlation coefficient is invariant to amplitude shifting and stretching and also ranges from -1 to 1 \cite{siming_thesis}. 

Dynamic time warping (DTW) identifies the path between two vectors of the lowest cumulative Euclidean distance by shifting the time axis. DTW is invariant to local and global time shifting and stretching \cite{Kong}. The DTW algorithm used in this study is only invariant to time shifting. DTW can range from 0 to $\infty$.

\begin{figure}[tbh]
    \centering
    \begin{subfigure}[t]{0.22\textwidth}
        \centering
        \includegraphics[height=1.3in]{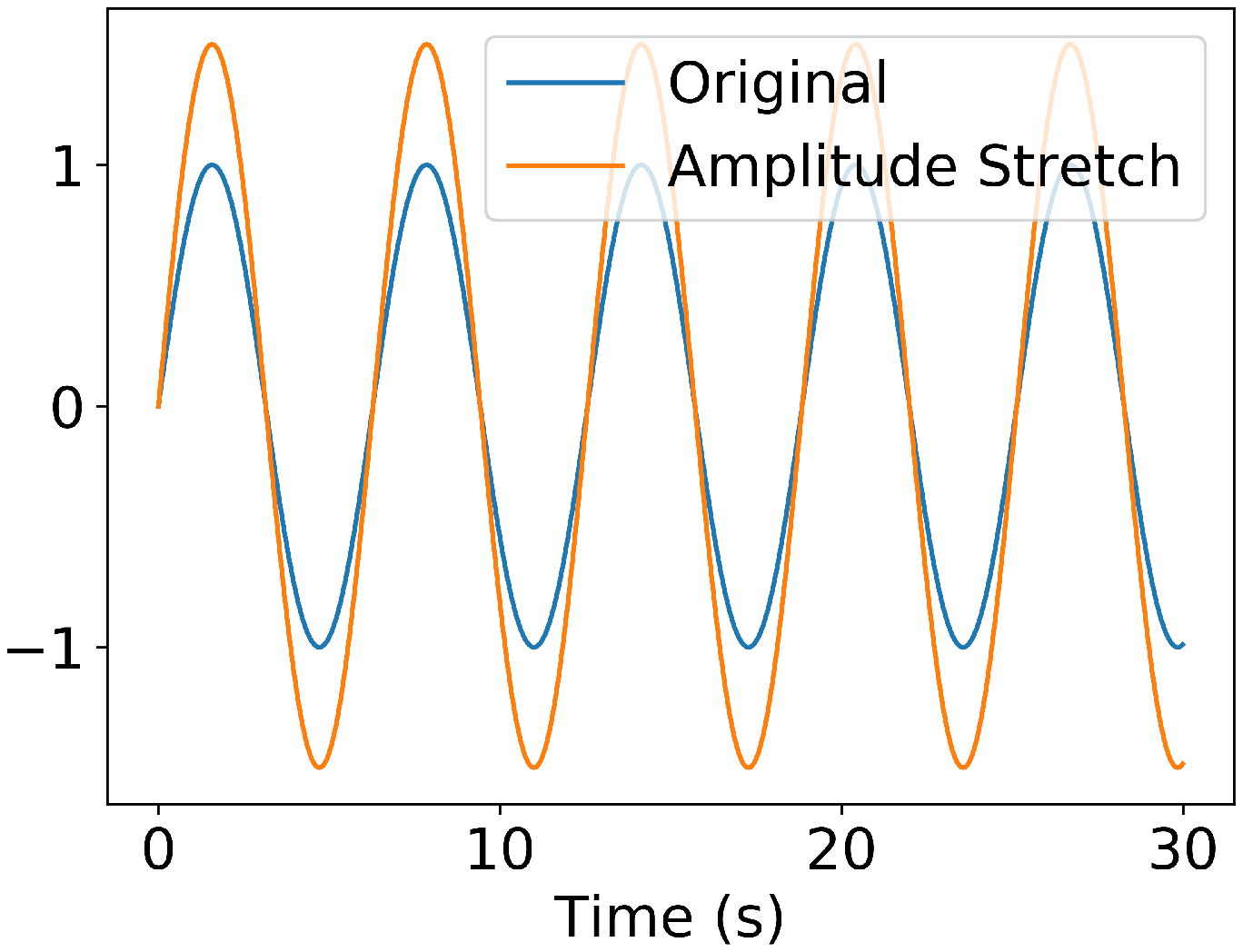}
        \caption{Amplitude stretch}
    \end{subfigure}
    ~ 
    \begin{subfigure}[t]{0.22\textwidth}
        \centering
        \includegraphics[height=1.3in]{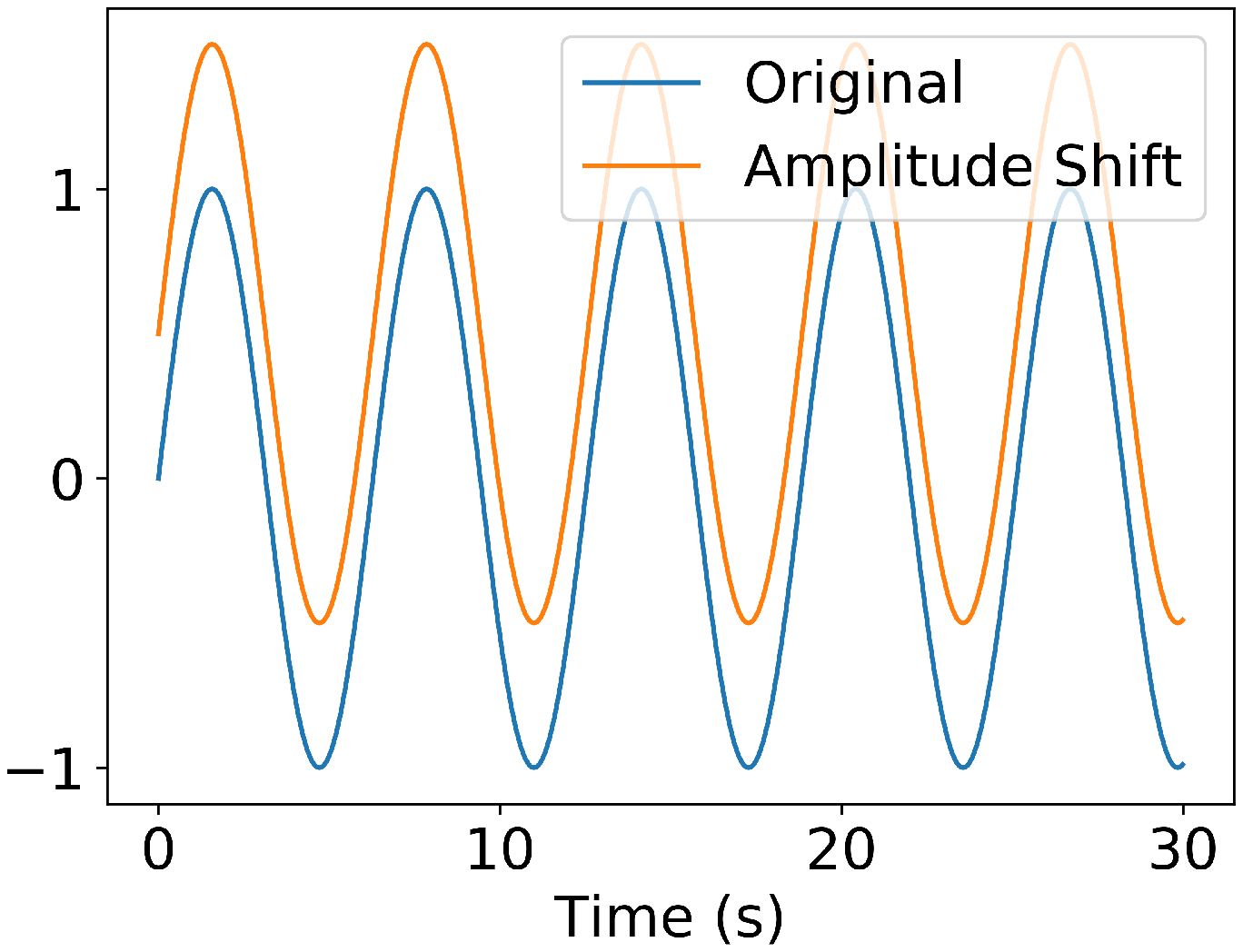}
        \caption{Amplitude shift}
    \end{subfigure}
    \begin{subfigure}[t]{0.22\textwidth}
        \centering
        \includegraphics[height=1.3in]{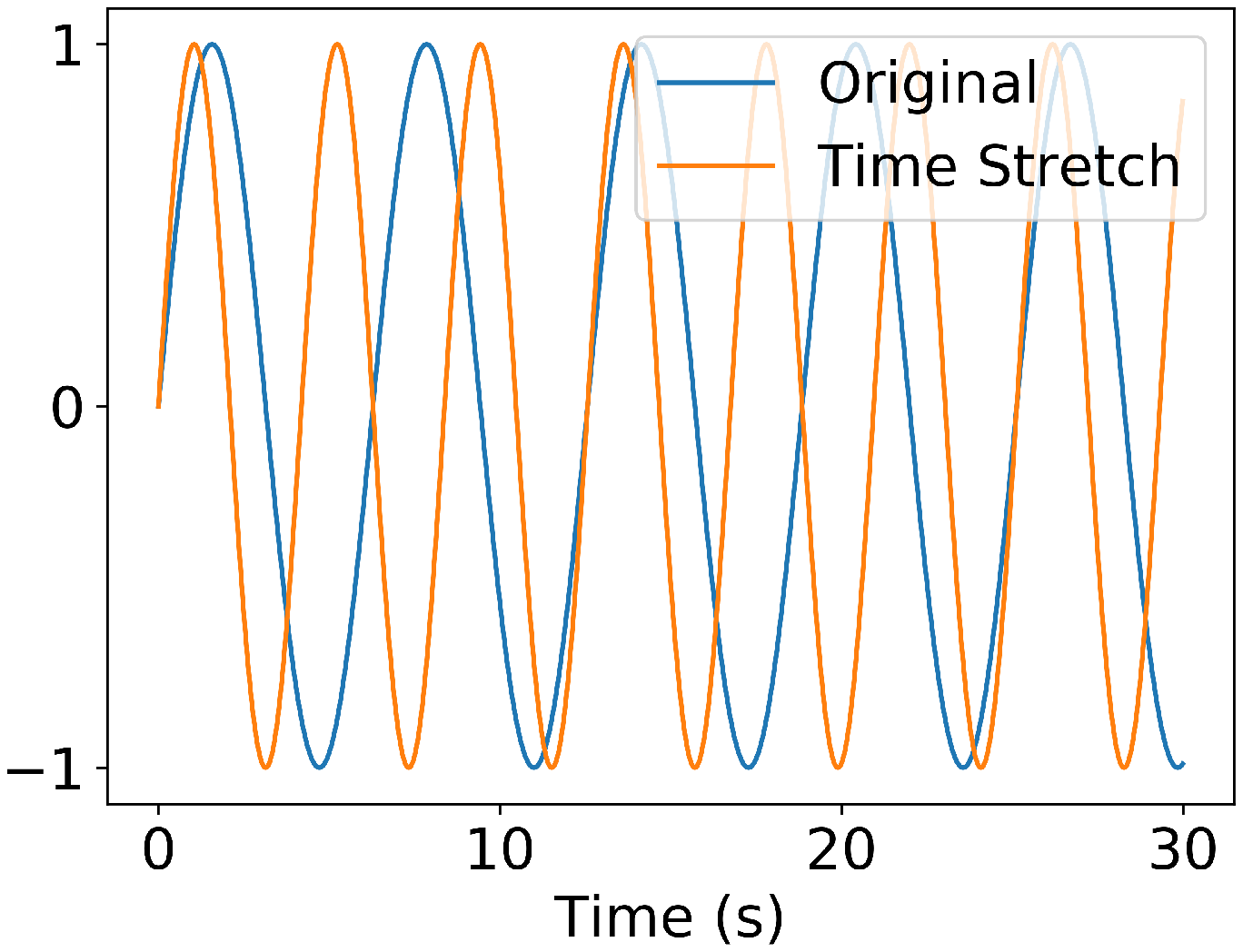}
        \caption{Time stretch}
    \end{subfigure}
    \begin{subfigure}[t]{0.22\textwidth}
        \centering
        \includegraphics[height=1.3in]{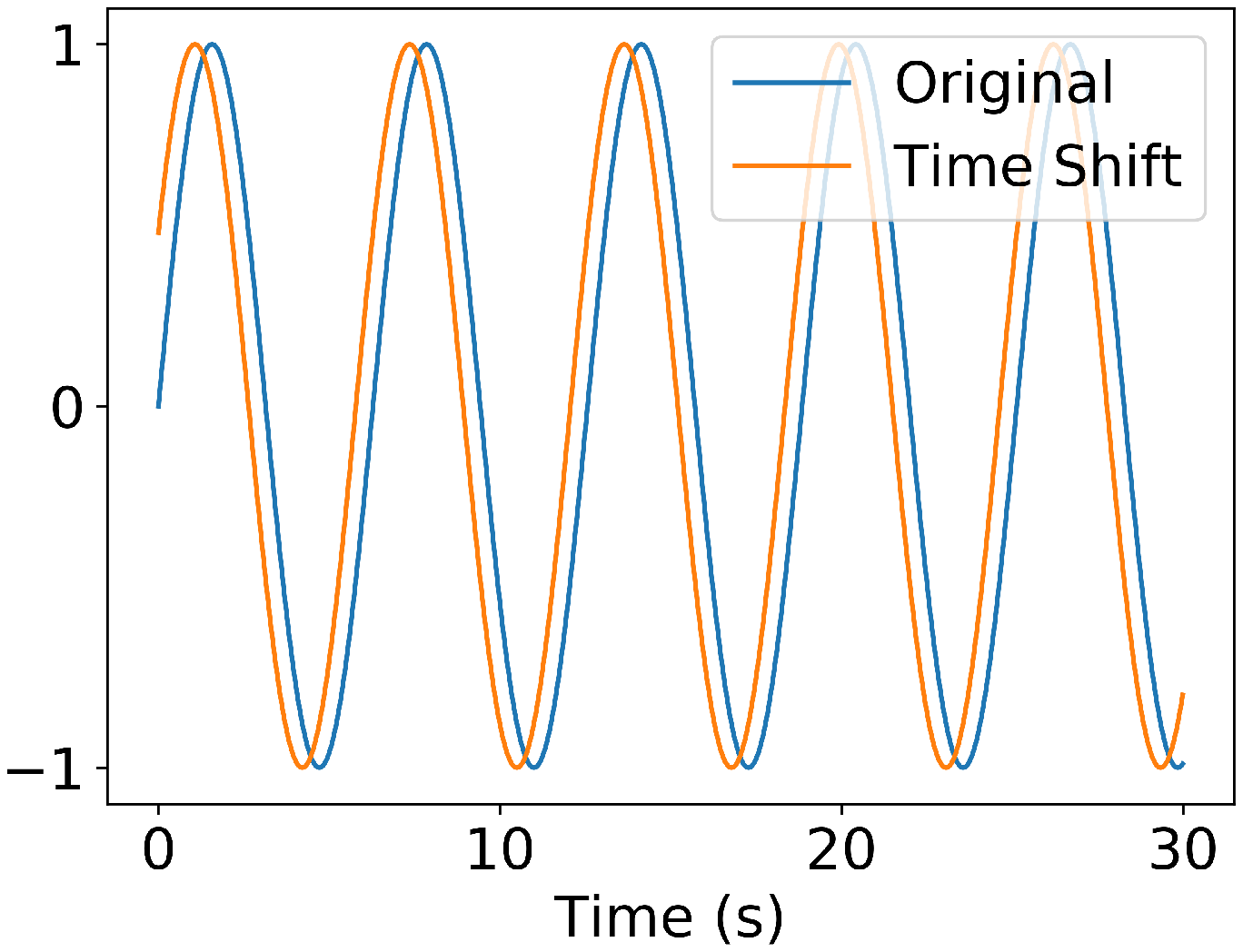}
        \caption{Time shift}
    \end{subfigure}
    \caption{Example time series with amplitude and time shift and stretch}
    \label{example_plots}
\end{figure}

\begin{figure}[thb]
    \centering
	\includegraphics[clip,width=1\linewidth]{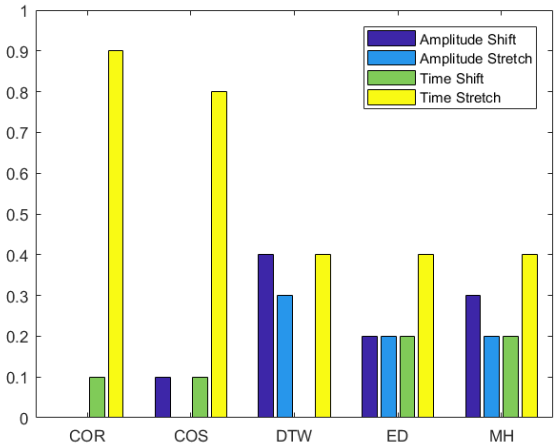}
    \caption{Comparison of similarity measures}
    \label{example_comparison}
\end{figure}

Figure \ref{example_plots} and \ref{example_comparison} show how amplitude and time shifting and stretching affect the error produced by similarity measures. The time series plots in Figure \ref{example_plots} show a sine wave with corresponding amplitude or time shift or stretch. The similarity measures calculate the difference between each of the time series subplots. The error generated for each similarity measure is normalized for comparison. The error is normalized separately for each similarity measure, so the sum of the error from the amplitude and time shift and stretch sums to one. Figure \ref{example_comparison} compares the error results from each of the subplot scenarios. 

The results in Figure \ref{example_comparison} demonstrate the abilities of each similarity measure. The similarity measures are denoted as: Euclidean distance (ED), Manhattan distance (MH), dynamic time warping (DTW), cosine distance (COS), and correlation coefficient (COR). Correlation coefficient has negligible error produced with both amplitude shift and stretch. Cosine distance has negligible error with amplitude stretch. Dynamic time warping has negligible error with time shift. These results provide an example of what can be expected when they are used with simulation or PMU time series data.

\section{System Level Experiment Methodology}

The system level experiment is setup to determine whether system response error can determine the total system accuracy. This addresses the question: is it possible to determine if any models or the approximate percentage of models in the system are inaccurate and need to be updated, with out needing to test at each individual bus? This is determined by calculating the correlation between system accuracy, as defined in Equation \ref{system_accuracy}, and system response error described below. 

This experiment is performed within the RTS96 test system \cite{RTS96,Lassetter}, using Siemens PSS/E software. Fourteen CLMs are randomly placed on loads in the system enhancing the RTS96 case to create a load model benchmark system. The remaining 37 loads are modeled with the static ZIP load model. Test systems are generated by replacing some ZIP load models from the benchmark system with CLM in the test system and some CLM in the benchmark system to ZIP load models in the test system. Switching load models creates "inaccurate" and "accurate" load models as a method to change the accuracy of the system. The "inaccurate" load models are those in the test system that are different from the benchmark system. The buses with the same load models in the test system and benchmark system are "accurate" load models. Switching these load models will also create difference responses, as described in Section I.  

A hundred of benchmark and test systems are created using the randomized placement of CLMs, based on a uniform random distribution, to reduce the sensitivity of the results to location of the CLM in the system. The percentage of buses in the test system with accurate load models is called the system accuracy. System accuracy is defined in Equation \ref{system_accuracy} and is also used in the Bus Level Experiment. 

\begin{dmath}\label{system_accuracy}
    \centering
    \textnormal{accuracy}_{\textnormal{system}} =  \frac{\textnormal{Buses\ with\ accurate\ load\ models}}{\textnormal{total\ number\ of\ buses\ with\ loads}}
\end{dmath}

An example benchmark and test system pair at 50\% system accuracy will have half of the CLMs removed from the benchmark system. The removed CLMs will be replaced with ZIP load models. System accuracy quantifies how many dynamic load models in the system are accurate. Accurate dynamic load models in the test systems are those models which are the same as those in the benchmark system. 

A bus fault is used to create a dynamic response in the system. Over a hundred simulations are performed where the location of the fault is randomized to reduce the sensitivity of fault location in comparison to CLM location. The bus fault is performed by applying a three-phase to ground fault with a duration of 0.1 s. During this fault, there is an impedance change at the bus fault causing the voltage to drop at the bus and a change in power flows throughout the system. The fault is cleared 0.1 seconds after it is created, and the power flows returns to a steady-state. 

The output captured from the simulations are voltage magnitude, voltage angle, and frequency from all of the load buses, and line flow active power and reactive power. The output from the benchmark system is compared to the test systems using the similarity measures outlined in Section \ref{similarity_section}. The response error generated by DTW, cosine distance, and correlation coefficient are a single measure for the entire time span of each output at each bus. The response error from Manhattan and Euclidean distance is generated at every time step in the time span. The error at each time step is then summed across the time span to create a single response error similar to the other similarity measures. The generation of response error for Manhattan and Euclidean distance is shown by Equation \ref{response_accuracy}. 

\begin{equation}\label{response_accuracy}
    \centering
    \textnormal{error}_{\textnormal{response}} = \sum_{t=1}^Ts[t]
\end{equation}

Similar to response error, system response error is calculated from the difference between the output of buses between the benchmark and test systems. However, system response error is a single metric which is the sum of all the response errors from each bus.

Three time spans are tested: 3 seconds, 10 seconds, and 30 seconds. The disturbance occurs at 0.1 seconds and cleared at 0.2 seconds for all the scenarios. These time spans are chosen to test the sensitivity to the transient event occurring in the first 3 seconds, and sensitivity to the dynamic responses out to 30 seconds. 

The Pearson correlation coefficient is calculated between system accuracy and system response error using the student t-test, to determine the relationship between the two. The student t-test is a statistical test to determine if two groups of results being compared have means which are statistically different. The output of the Pearson correlation coefficient is the r and p-value. The r-value denotes the direction and strength of the relationship. R-values range from -1 to 1, where -1 to -0.5 signifies a strong negative relationship and 0.5 to 1 signifies a strong positive relationship between the groups. For this experiment, a strong negative relationship implies that as the system accuracy increases the system response error decreases. This is the relationship typically assumed by those performing measurement-based techniques. The p-value is the value which determines if the two results are different. A p-value of less than 0.05 signifies a statistically significant difference between the two groups of results being compared. Therefore, a p-value less than 0.05 signifies a statistically significant relationship quantified by the r-value.

\section{System Level Experiment Results}

In this section, the correlation between response and system accuracy is calculated to evaluate the ability of various time spans, output types, and similarity measures to predict system accuracy as used in measurement-based techniques. 

\begin{figure}[thb]
    \centering
	\includegraphics[clip,width=.9\linewidth]{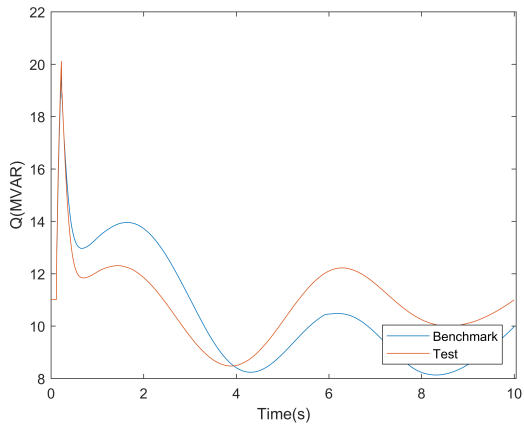} 
    \caption{Reactive Power Time Series Plot of Low System Accuracy and High Response Error with Generator Outage}
    \label{high_accuracy}
\end{figure}

\begin{figure}[thb]
    \centering
	\includegraphics[clip,width=.9\linewidth]{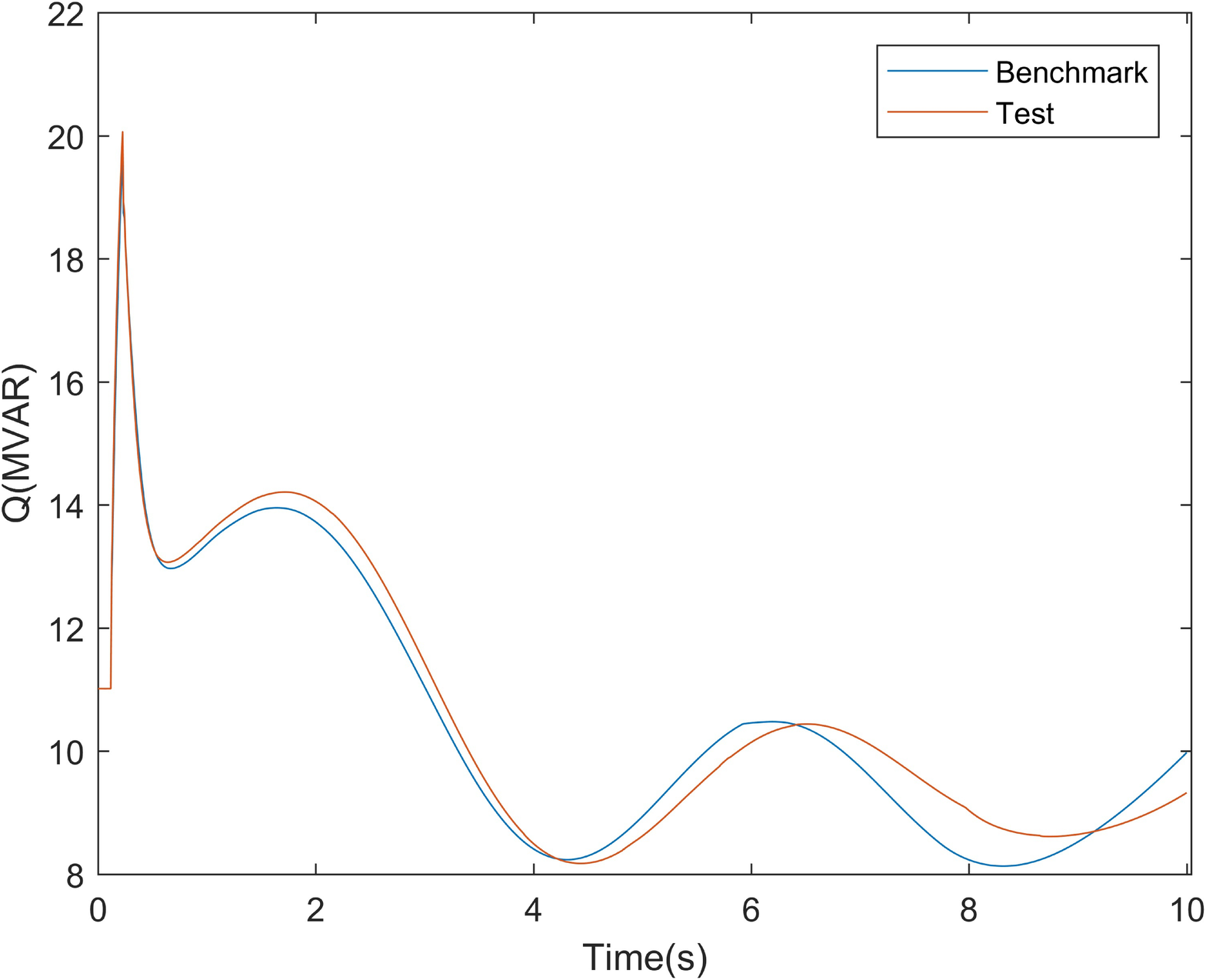} 
    \caption{Reactive Power Time Series Plot of High System Accuracy and Low Response Error with Generator Outage}
    \label{low_accuracy}
\end{figure}

An example outputs from these results is visualized in Figures \ref{high_accuracy} and \ref{low_accuracy}. The plots compare the reactive power times series data from a bus in the benchmark system and test systems at two levels of system accuracy in a system undergoing a bus fault at the same bus. Figure \ref{low_accuracy} shows the benchmark and test system responses with low system accuracy, 8\%. Figure \ref{high_accuracy} shows the responses with high system accuracy, 92\%.

The response from the high system accuracy test system has a better curve fit to the benchmark system than the low system accuracy test system. This visual comparison confirms that with an appropriate similarity measure the response error should decrease as system accuracy increases. 

\begin{figure}[tbh]
    \centering
    \begin{subfigure}[t]{0.5\textwidth}
        \centering
        \includegraphics[height=1.6in]{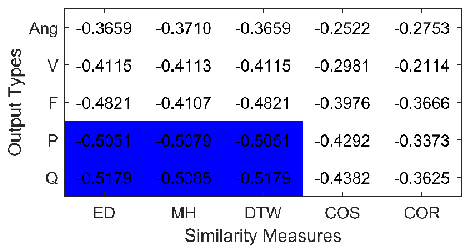}
        \caption{3 second time span}
    \end{subfigure}
    ~ 
    \begin{subfigure}[t]{0.5\textwidth}
        \centering
        \includegraphics[height=1.6in]{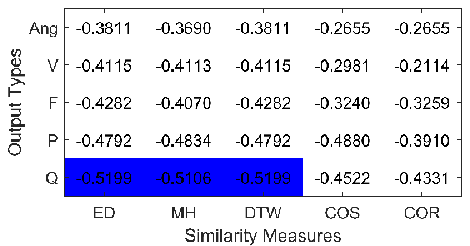}
        \caption{10 second time span}
    \end{subfigure}
    \begin{subfigure}[t]{0.5\textwidth}
        \centering
        \includegraphics[height=1.6in]{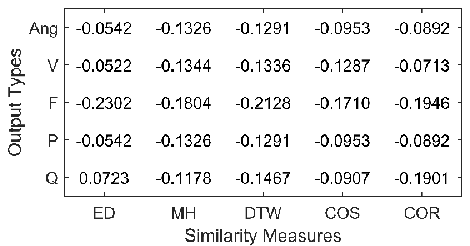}
        \caption{30 second time span}
    \end{subfigure}
    \caption{Bus fault R-values for system level experiment for time spans: a) 3 seconds, b) 10 seconds, c) 30 seconds}
    \label{R_LO_3}
\end{figure}

The results from all the simulations determining correlation between system accuracy and response error as grouped by the metrics used are shown as r-values in Figure \ref{R_LO_3}. R-values of less than -0.5 are highlighted in orange to show they represent a strong relationship. R-values greater than -0.5, which do not have a strong relationship, are in white. All resulting p-values are found to be lower than 0.05, meaning all r-value relationships are statistically significant. The similarity measures listed in the plots use the same abbreviations as in Figure \ref{example_comparison}. The output types listed in the plots are abbreviated with: voltage angle (ANG), voltage magnitude (V), frequency (F), line active power flow (P), and line reactive power flow (Q). 

Out of the 75 combinations of metrics tested in this experiment, only 12\% yielded statistically significant differences. Considering the visual verification that indeed response error should decrease as system accuracy increases from Figures \ref{high_accuracy} and \ref{low_accuracy}, the lack of strong negative correlations seen in Figure \ref{R_LO_3} are concerning. Only the three and ten-second time span simulations have strong correlation relationships, none of the thirty-second scenarios have strong relationships. During a thirty-second simulation, the last ten to thirty seconds of the output response will flatten to a steady-state value. Therefore, in a thirty-second simulation there are many error data points that might contain flat steady-state responses limiting curve fitting opportunities and reducing a correlation relationship. This can explain why none of the thirty-second scenarios have strong relationships. 

\begin{figure}[thb]
    \centering
	\includegraphics[clip,width=1\linewidth]{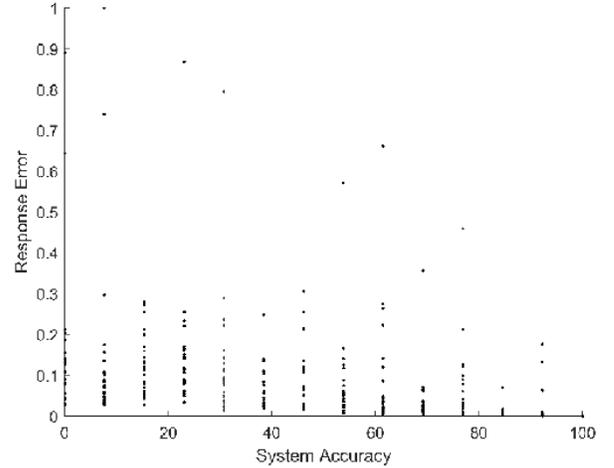}
    \caption{R-value distribution}
    \label{distribution}
\end{figure}

The distribution of the r-values from the overall strongest correlation relationship, with an r-value of -0.5199, is examined to further investigate the correlation results. Figure \ref{distribution} visualizes the distribution of the response error for this r-value at the tested levels of system accuracy.  

The response error in figure \ref{distribution} is normalized for a clearer comparison. A general negative correlation is seen, where there is lower response error at higher system accuracy. However, there are several outliers in the data preventing a stronger overall correlation, particularly between system accuracy levels 0\% and 70\%. This suggests at lower system accuracy levels the correlation is not as high as in the overall distribution. To test this, the correlation between system accuracy ranges is calculated to highlight where the weakest correlation  regions exist. Table \ref{correlation_ranges} outlines the correlation at the following system accuracy ranges.

\begin{table}[thb]
\centering
\caption{\label{correlation_ranges} Correlation within system accuracy ranges}
\begin{tabular}{p{1.5cm} p{1.5cm} p{1.5cm} p{1.5cm}}
\hline \\
 0-30\% & 38\%-54\% & 62\%-77\% & 84\%-100\%  
 \\
 \hline
 \\ 
 -0.0632 & -0.2964 & 0.0322 & -0.4505 
 \\
\hline
\end{tabular}
\end{table}

Seen in Table \ref{correlation_ranges}, the correlation is greatly degraded at the low levels of the system accuracy ranges, even reversing the r-value relationship from negative to positive between levels 62\% and 77\%. An ideal scenario would have a constant strong negative correlation through all system accuracy levels. This highlights a potential low effectiveness of measurement-based techniques using these testing conditions at low system accuracy levels. Overall, the results from this experiment highlight the lack of correlation between response error and system accuracy across all metrics. 

The application of the system level experiment is to use any of the metrics combinations that showed strong negative relationships in a measurement-based optimization program. Such an optimization program could change the dynamic load models in the system to reduce system response error in order to improve system accuracy. However, in order for such an optimization program to successfully improve system accuracy, there needs to be a strong negative correlation between system accuracy and system response error. Additionally, even with an overall strong negative correlation, Table \ref{correlation_ranges} shows that such an optimization program may determine a local minimum at a lower accuracy level to be the global minimum due to the lower correlation relationship strength found at lower accuracy levels. 

These results identify the need for measurement-based techniques, and potentially other power systems time series data curve fitting techniques, to evaluate the assumption that the system response error is correlated to the system accuracy. It cannot be assumed measurement-based techniques using similarity measures yield meaningful results. Any optimization or other estimation technique using the reduction of system response error will not yield accurate results of findings without a strong correlation between system response error and system accuracy.

\section{Bus Level Experiment Methodology}

The bus level experiment is setup to determine whether response error from an individual load bus can indicate if a load model being used at the bus is accurate. In comparison to the system level experiment which looked at system wide model accuracy, this experiment looks at model accuracy at the bus level. The results of this experiment are the p-values from the student t-test, indicating whether there is a statistical difference between the response error from buses with accurate and inaccurate load models. The p-value is the value which determines if the two results are different. A p-value of less than 0.05 signifies a statistically significant difference between the two groups of results being compared.

The same system and system setup are used in this experiment as in the system level experiment. This experiment excludes comparing the output from line flow active power and reactive power with the previously used outputs of frequency, voltage angle, and voltage magnitude of the buses. In this experiment the simulations are performed at various levels of system accuracy to reduce the sensitivity of the results to the system accuracy. By reducing the sensitivity of the results to fault placement and system accuracy, the results focus the correlation to between response error and load model accuracy. All other metrics remain the same as the system level experiment.

The response error from all the simulations are compared by output type, time span, and similarity measure, and binned into groups of buses with accurate load models and buses with inaccurate load models. A t-test is performed on the binned response error to determine if there is a statistically significant difference between the error from buses with accurate load models and buses with inaccurate load models. The results of this experiment are the p-values from the response error separated by disturbance scenario, output type, time span, and similarity measure.

\section{Bus Level Experiment Results}

\begin{figure}[tbh]
    \centering
    \begin{subfigure}[t]{0.5\textwidth}
        \centering
        \includegraphics[height=1.6in]{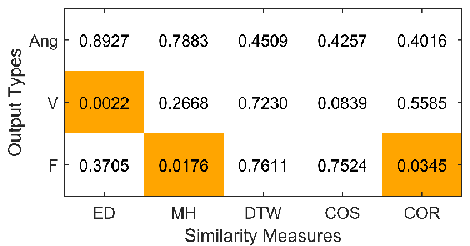}
        \caption{3 second time span}
    \end{subfigure}
    ~ 
    \begin{subfigure}[t]{0.5\textwidth}
        \centering
        \includegraphics[height=1.6in]{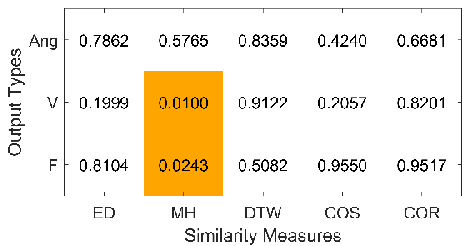}
        \caption{10 second time span}
    \end{subfigure}
    \begin{subfigure}[t]{0.5\textwidth}
        \centering
        \includegraphics[height=1.6in]{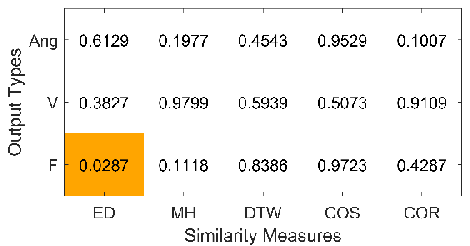}
        \caption{30 second time span}
    \end{subfigure}
    \caption{Bus fault P-values for bus level experiment for time spans: a) 3 seconds, b) 10 seconds, c) 30 seconds}
    \label{Bus_level_BF}
\end{figure}

The bus level experiment tests whether there is a statistical difference between the response error at individual buses with the accuracy of the load models at the buses. The p-values are calculated using response error from the output types, time spans, and similarity measures. Figure \ref{Bus_level_BF} shows these p-values. 

Less than 15\% of the combinations of time span, output type, and similarity measure have significant p-values. It is noted that the combinations of metrics best used for this experimental setup are different than those in the system level experiment. This experiment highlights a serious concern for other experiments using measurement-based techniques. Only select combinations of metrics in this experiment yielded significant differences, and this same result is likely present with other measurement-based experiments whether they involve changing load models, changing load model parameters, or changes in other dynamic models.    

The direct application of this experiment is to use any of the disturbance type, output type, time span, and similarity measure combinations that showed significant p-values in a measurement-based machine learning technique to identify if a bus in the system need a load model updated or a different load model. There needs to be a significant difference between response errors from buses with poor fitting or inaccurate load models and those which are accurate for such a machine learning algorithms to give meaningful results, whether it be from simulation or PMU outputs. In this case, if the machine learning algorithm was using a combination of metrics that did not have a proven significant difference between response error from buses with inaccurate and accurate load models, the machine learning algorithm would be unable to accurately tell the difference between the groups, causing the results to be inaccurate.

The results from this experiment confirm the same conclusion from the system level experiment that there needs to be verification testing showing that the chosen measurement-based metrics used to calculate error will capture true differences between incorrect models and correct models. It cannot be assumed that any combination of metrics used in measurement-based techniques will yield meaningful results.

\section{Conclusion}

This paper investigates common metrics used in measurement-based dynamic load modeling techniques to generate response error. These metrics include similarity measures, output types, and simulation time spans. The correlation between response error and accuracy is evaluated by comparing the system accuracy to system response error with the system level experiment and load model accuracy to bus response error with the bus level experiment. Both experiments demonstrated there is a lack of combinations of metrics that deliver significant findings. It is noted that the combinations of metrics best used in the bus level experiment are different than those in the system level experiment. This same result is likely to be found with other measurement-based experiments whether they involve changing load models, changing load model parameters, or changes in other dynamic models. These experiments expose a significant concern for measurement-based technique validity. This study raises awareness of the importance of careful selection and validation of similarity measures and response output metrics used, noting that naive or untested selection of metrics can deliver inaccurate and meaningless results.

These results implicate that optimization or machine learning algorithms that use measurement-based techniques without validating their metrics to ensure correlation between error and accuracy may not generate accurate or meaningful results. These methods to determine the effectiveness of the use of these common metrics are specific to these experiments of model accuracy. Future work can expand these methods to dynamic model parameter tuning experiments.

\bibliographystyle{ieeetr}
\bibliography{ref}

\begin{IEEEbiography}[{\includegraphics[width=1in,height=1.25in,clip,keepaspectratio]{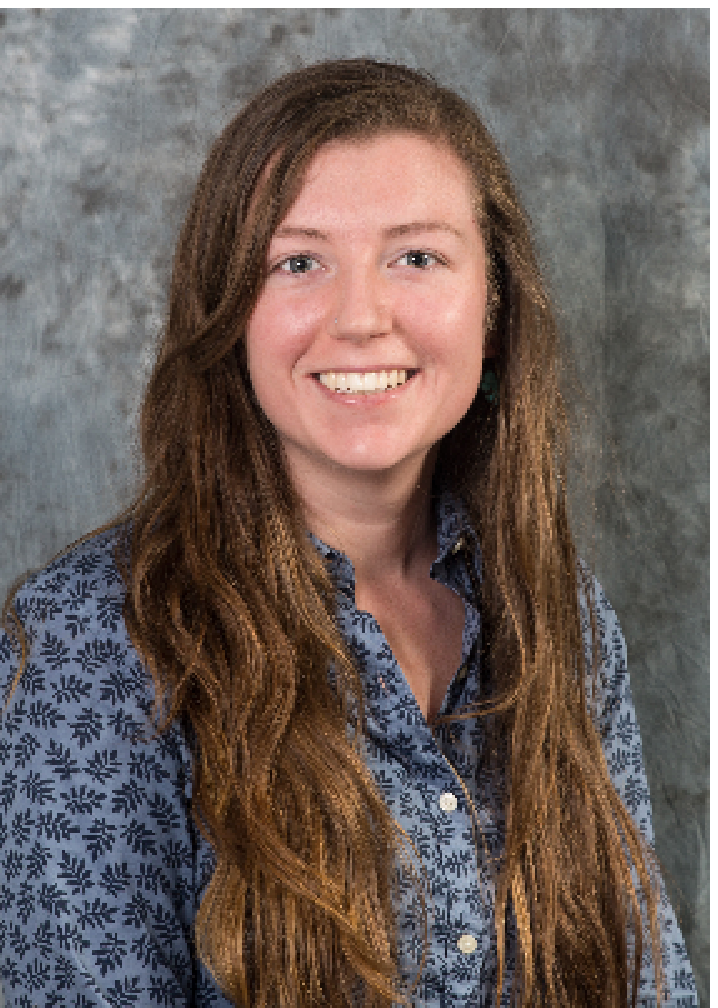}}]{Phylicia Cicilio}
(S'15) received the B.S. degree in chemical engineering in 2013 from the University of New Hampshire, Durham, NH, USA. She received the M.S. degree in electrical and computer engineering in 2017 from Oregon State University, Corvallis, OR, USA, where she is currently working toward the Ph.D. degree in electrical and computer engineering. 

She is currently a Graduate Fellow at Idaho National Laboratory, Idaho Falls, ID, USA. Her research interests included power system reliability, dynamic modeling, and rural electrification.  

\end{IEEEbiography}

\begin{IEEEbiography}[{\includegraphics[width=1in,height=1.25in,clip,keepaspectratio]{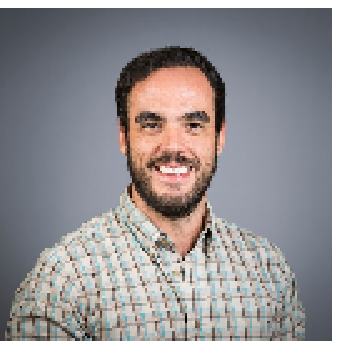}}]{Eduardo Cotilla-Sanchez}
(S'08-M-12-SM-19) received the M.S.~and Ph.D.~degrees in electrical engineering from the University of Vermont, Burlington, VT, USA, in 2009 and 2012, respectively.

He is currently an Associate Professor in the School of Electrical Engineering and Computer Science, Oregon State University, Corvallis, OR, USA. His primary field of research is electrical infrastructure resilience and protection, in particular, the study of cascading outages.

Prof.~Cotilla-Sanchez is the Vice-Chair of the IEEE Working Group on Cascading Failures and President of the Society of Hispanic Professional Engineers Oregon Chapter. 

\end{IEEEbiography}

\end{document}